# Confocal structured illumination microscopy for super-resolution imaging: theory and numerical simulations

**Junzheng Peng[1], Jiahao Xian[1], Xi Lin[1], Manhong Yao[2], Shiping Li[1], Jingang Zhong[1,*]**

[1]*Department of Optoelectronic Engineering, Jinan University, Guangzhou 510632, China*

[2]*School of Optoelectronic Engineering, Guangdong Polytechnic Normal University, Guangzhou 510665, China*

* Corresponding author, E-mail: tzjg@jnu.edu.cn

**Abstract:** Super-resolution structured illumination microscopy (SR-SIM) is a widely used technique for enhancing the resolution of fluorescence imaging beyond the diffraction limit. Most existing SR-SIM methods rely on Moiré effect-based physical imaging models, which require estimating structured illumination parameters during image reconstruction. However, parameter estimation is prone to errors, often leading to artifacts in the reconstructed images. To address these limitations, we propose super-resolution confocal structured illumination microscopy (SR-CSIM). The physical model of SR-CSIM is based on confocal imaging principles, eliminating the need for structured illumination parameter estimation. We construct the SR-CSIM imaging theory. Numerical simulation results demonstrate that SR-CSIM achieves a resolution comparable to that of SR-SIM while reducing artifacts. This advancement has the potential to broaden the applicability of SIM, providing researchers with a more robust and versatile imaging tool.

## 1. Introduction

Structured illumination microscopy (SIM) [1] has emerged as a powerful tool in fluorescence imaging, particularly in super-resolution imaging [2,3] and optical sectioning imaging [4]. As a computational imaging method, super-resolution structured illumination microscopy (SR-SIM) utilizes time-varying structured illuminations and post-processing algorithms to reconstruct high-resolution images. The fidelity of the reconstructed image depends on the physical imaging system and the image reconstruction algorithm based on the physical imaging model.

The physical imaging model in existing SR-SIM is fundamentally based on the Moiré effect [5]. Through this phenomenon, the high-frequency information beyond the resolution limit of conventional microscopes is transferred into the detectable low-frequency region via spatial frequency mixing. Reconstruction algorithms in SR-SIM operate in the spatial frequency domain (Fourier space) to extract this high-frequency information embedded in the mixed-frequency signals. Once extracted, this high-frequency information is shifted back to its original frequency positions and integrated into an extended Fourier spectrum. Finally, through an inverse Fourier transform, a super-resolution image approximately twice the resolution of conventional wide-field microscopy is generated in the spatial domain (real space). However, this process necessitates the accurate estimation of the illumination structured light parameters, such as spatial frequency, initial phase, and modulation depth. Any inaccuracies in these estimations can introduce artifacts in the final image [6], and despite recent advancements in improved algorithms [7–11], the issue of artifacts persists as a limitation in SR-SIM [12,13].

In the realm of fluorescence microscopy, confocal scanning microscopy, pioneered by Marvin Minsky in 1957, represents a milestone [14]. It enables three-dimensional (3D) imaging and has become an essential tool across various scientific and medical fields [15]. Compared to SIM, confocal microscopy offers superior noise resistance and enhanced imaging depth due to the use of a pair of pinholes that effectively filter the background and scattered noise before image formation [16]. Nevertheless, its lateral resolution remains constrained by the sizes of the illumination spot and the detection pinhole [17]. Theoretically, reducing the pinhole size to achieve lateral super-resolution is possible [18], though this approach results in substantial signal loss, leading to images with poor signal-to-noise ratio (SNR). To mitigate this limitation, researchers have developed solutions such as replacing the point detector with an array detector, such as a camera, and implementing pixel reassignment algorithms [19,20] to reconstruct high SNR super-resolution images, giving rise to image scanning microscopy (ISM) [21,22]. Additionally, advanced techniques like stimulated emission depletion (STED) microscopy [23,24] and single-molecule localization microscopy (SMLM) [25–27] can be viewed as confocal imaging methods that reduce the illumination spot size to achieve super-resolution.

Recently, we introduced the concept of confocal imaging into the optical-sectioning structured illumination microscopy (OS-SIM) and proposed confocal structured illumination microscopy (CSIM) to enhance both the SNR and imaging depth of optical sectioning for fluorescent samples [28]. We termed this technique as optical sectioning CSIM (OS-CSIM). While sharing the same optical imaging system as OS-SIM, OS-CSIM distinguishes itself through its use of a confocal imaging physical model for image reconstruction. Specifically, it reconstructs dual images for each camera pixel based on the principles of dual imaging and single-pixel imaging [29–32]. By applying the confocal imaging principle, OS-CSIM isolates the intensity values of the object point conjugated to each camera pixel from the dual images, thereby constructing a confocal image. This method demonstrates improved SNR and imaging depth compared to traditional OS-SIM. However, its lateral resolution remains comparable to that of conventional wide-field microscopy.

To further advance the capabilities of structured illumination microscopy, we established a theoretical framework for super-resolution CSIM (SR-CSIM). Unlike traditional SR-SIM, which relies on the Moiré effect for image reconstruction, SR-CSIM is grounded in the principle of confocal imaging. Departing from conventional confocal microscopy that employs point-by-point illumination, SR-CSIM achieves confocal imaging under wide-field structured illumination generated via a two-beam interference. The workflow involves treating each camera pixel as a single-pixel detector to reconstruct an image that separates the signals from different object points recorded by each pixel. Subsequently, SR-CSIM utilizes confocal imaging principles to extract the conjugate signals, thereby constructing a super-resolution image. A key contribution of SR-CSIM is the elimination of the need for structured illumination parameter estimation, thereby avoiding artifacts that commonly arise from SR-SIM due to inaccuracies in parameter estimation. This breakthrough positions SR-CSIM as a robust and versatile imaging tool, expanding the applicability of SIM in various imaging scenarios.

## 2. Principle and method

### 2.1 Principle

The imaging system of SR-CSIM, as illustrated in Fig. 1, shares the same hardware configuration as the traditional SR-SIM setup. The key contribution of SR-CSIM lies in the image reconstruction algorithm, which is based on the principles of dual imaging and confocal imaging. The laser beam emitted from the light source (LS) undergoes initial expansion and collimation before being directed onto the modulation plane of the spatial light modulator (SLM). A digital grating preloaded on the SLM diffracts the laser beam. After passing through Lens (L1), the diffracted light converges at different positions on the focal plane. A mask is employed to selectively transmit the +1 order and the -1 order diffracted beams, while effectively blocking the 0 order and other high-order diffracted lights. These filtered beams then travel through Lens 2 (L2), an illuminating tube lens (ITL), a dichroic mirror (DM), and an objective lens (OL), ultimately interfering in the object space to generate cosine-structured light. This structured light illuminates the fluorescent sample (S), exciting fluorescence emission. The emitted fluorescence is collected by the objective lens, passes through the DM and the detecting tube lens (DTL), and is finally focused onto the photosensitive plane of the camera (CAM) for recording.

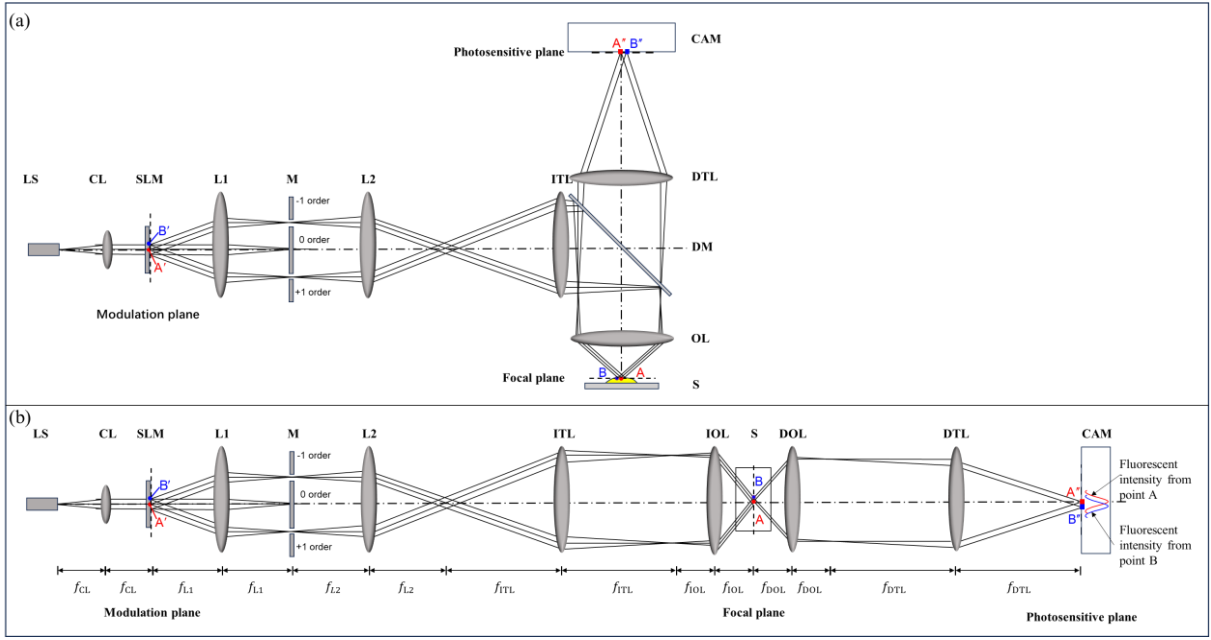

Fig. 1(a) Schematic diagram of the SR-CSIM imaging system; (b) the equivalent optical path of Fig. (a). (LS: light source; CL: collimated lens; SLM: Spatial light modulator; M: mask; L2: lens 2; DM: Dichroic mirror; ITL: illuminating tube lens; OL: objective lens; DTL: detecting tube lens; S: sample; CAM: camera.)

The equivalent optical path of the SR-CSIM imaging system is illustrated in Fig. 1(b). The modulation plane of the SLM, the object plane, and the camera's photosensitive plane are mutually conjugated. Points A′ and B′ on the modulation plane of the SLM are conjugated with the points A and B on the sample, respectively. They are also conjugated with the points A″ and B″ on the photosensitive plane of the camera. However, due to the inherent diffraction of light, the image of points A and B do not form ideal sharp points on the camera's photosensitive plane. Instead, they appear as Airy disks. As a result, each camera pixel, such as A″, records signals from multiple object points, including A and B. Similarly, the neighboring camera pixel B″ records the signals from B and A. This overlap of signals from adjacent object points limits the resolution of the image captured by the camera.

To address this limitation and improve imaging resolution, SR-CSIM employs three sets of phase-shifting cosine-structured light patterns (as shown in Fig. 2(a)) to illuminate the sample (Fig. 2(b)). These patterns have distinct spatial frequencies of $(f_{x_o}^1, f_{y_o}^1)$, $(f_{x_o}^2, f_{y_o}^2)$, and $(f_{x_o}^3, f_{y_o}^3)$. The camera captures the corresponding images of the sample under each illumination pattern (Fig. 2(c)). Since the object points A and B are optically conjugate with the SLM pixels A′ and B′, their spatial information is uniquely encoded by the structured illumination patterns. By treating each camera pixel (such as $A''$) as a single-pixel detector, a one-dimensional (1D) intensity sequence is recorded across the phase-shifting structured illumination (Fig. 2(d1)). Applying the single-pixel imaging method, a dual image is reconstructed by computationally separating the signals from object points A and B (Fig. 2(d2)) [29].

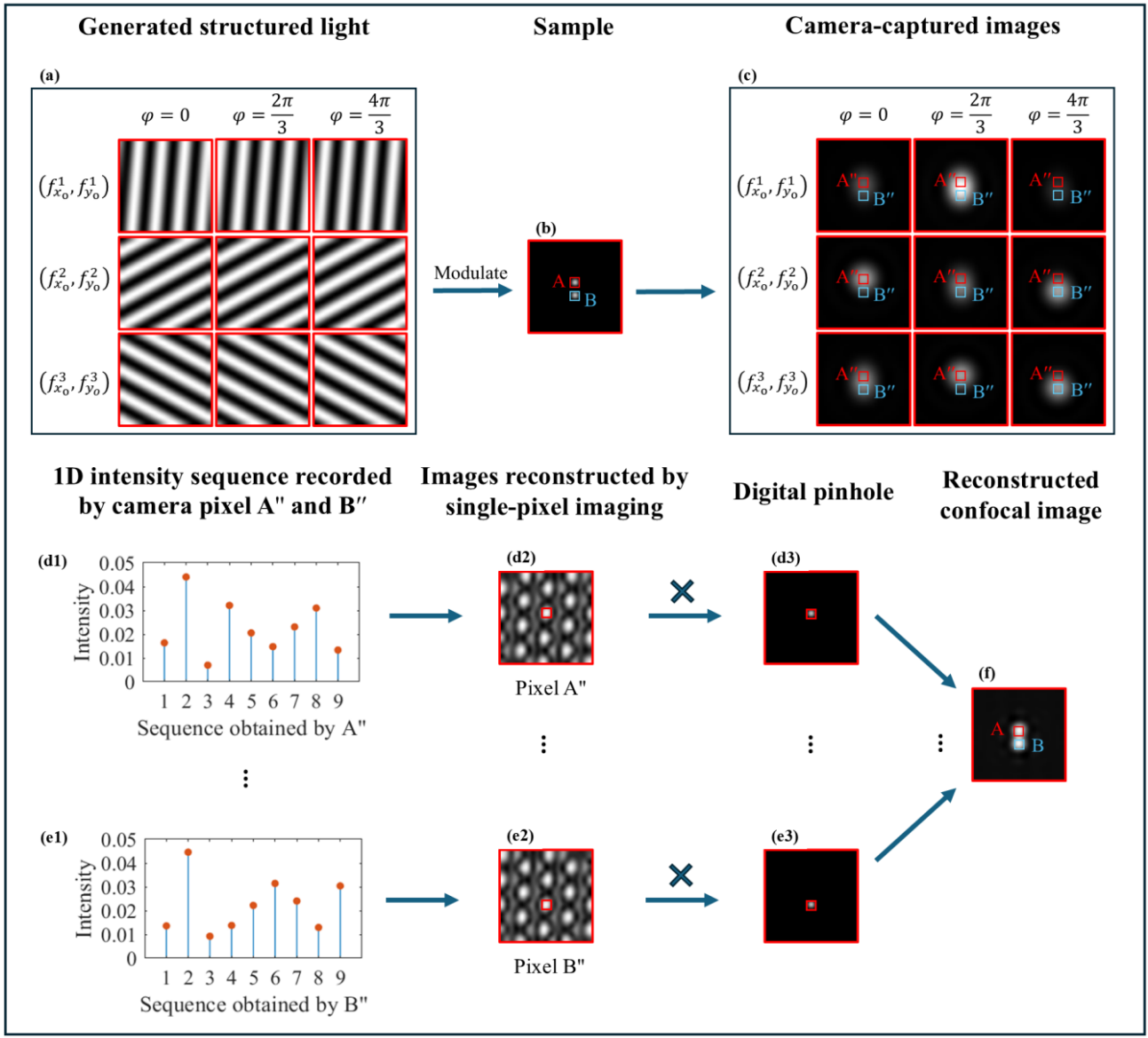

Fig. 2 Principle of SR-CSIM. (a) Structured light patterns generated on the object space; (b) measured sample; (c) camera-captured images; (d1-e1) 1D intensity sequences recorded by the pixel points $A''$ and $B''$; (d2-e2) images reconstructed by single-pixel imaging; (d3-e3) digital pinholes generated based on the conjugate relationship between the SLM and camera; (f) reconstructed super-resolution confocal image.

To further isolate the fluorescence signals from object point A, a digital pinhole is computationally generated (Fig. 2(d3)) based on the pre-calibrated conjugate relationship between the camera pixel point $A''$ and the SLM pixel point $A'$. This relationship can be calibrated using the method described in [28]. Utilizing the digital pinhole, the fluorescent signal emitted from object point A, which is optically conjugated to $A''$, can be selectively extracted from the dual image. Similarly, by treating the pixel point $B''$ as a single-pixel detector and applying the same methodology (as demonstrated in Figs. 2(e1-e3)), the fluorescent signal from object point B, conjugated to the pixel point $B''$, can also be extracted. This process is iteratively applied to all camera pixels to extract signals from their respective conjugate object points. Finally, the extracted signals are spatially reassembled according to the coordinates of the camera's pixel grid (Fig. 2(f)), reconstructing a confocal image.

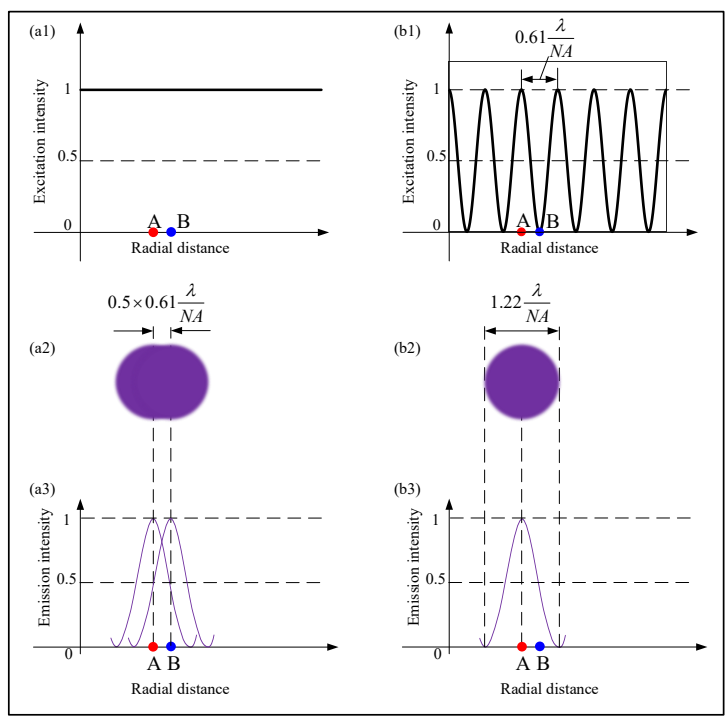

Fig. 3 Explanation of how SR-CSIM achieves super-resolution imaging. (a1) Illumination of object points A and B with a wide-field uniform light beam; (a2) fluorescent emission under uniform illumination; (a3) 1D intensity curve of Fig. (a2); (b1) illumination of object points A and B with wide-field structured light beam; (a2) fluorescent emission under structured illumination; (a3) 1D intensity curve of Fig. (b2)

The resolution of the reconstructed confocal image is fundamentally determined by the resolution of the single-pixel reconstructed dual image. According to the principle of single-pixel imaging, the resolution of the dual image is determined by the maximum spatial frequency of the structured illumination patterns employed. As illustrated in Fig. 3, consider two object points A and B separated by a distance $d = 0.5 \times 0.61\lambda/NA$, where $\lambda$ represents the emission wavelength of the fluorescence and $NA$ denotes the numerical aperture of the objective lens. Under wide-field uniform illumination (Fig. 3(a1)), both points A and B are simultaneously excited, resulting in overlapping fluorescence signals (Figs. 3(a2-a3)). According to the Rayleigh Criterion, these signals cannot be resolved under such conditions. In contrast, when the sample is illuminated with a sinusoidal intensity-modulated structured light pattern with a period of $0.61\lambda/NA$ (Fig. 3(b1)), object point A is selectively excited, while point B remains unexcited (Figs. 3(b2-b3)). By integrating multi-step phase-shifting algorithms (requiring $\geq 3$ steps) with single-pixel imaging techniques, the fluorescence signals from A and B are successfully separated. After this separation, the confocal imaging principle is applied to extract signals from conjugate object points based on the pre-calibrated SLM-camera conjugate relationship, enabling the reconstruction of a super-resolution confocal image.

A key distinction from traditional SR-SIM lies in the elimination of structured illumination parameter estimation. In SR-CSIM, the reconstructed dual image inherently reflects the pixel coordinates of the SLM, as dictated by the principles of single-pixel imaging [33]. Since the parameters of the structured light are predefined within the SLM's coordinate system, no parameter estimation is required. This fundamental distinction eliminates artifacts typically caused by parameter estimation errors in SR-SIM, ensuring artifact-free super-resolution imaging with enhanced fidelity.

## 2.2 Image reconstruction method

SR-CSIM first loads structural patterns onto the SLM to diffract the laser beam and filters the +1 and -1 order diffracted beams to interfere in the object space, generating cosine structured patterns. The intensity distribution of the cosine structured patterns generated on the object plane can be expressed as follows:

$$P(x_o, y_o; f_{x_o}, f_{y_o}; \phi_i) = a_o + b_o \cos(2\pi f_{x_o} x_o + 2\pi f_{y_o} y_o + \phi_i + \varphi_0), \tag{1}$$

where $(x_o, y_o)$ represents the spatial coordinate on the object plane, $a_o$ represents the background intensity, $b_o$ represents contrast, $(f_{x_o}, f_{y_o})$ represents the spatial frequency, $\varphi_0$ represents the initial phase for the spatial coordinate $(0,0)$, $\phi_i = i2\pi/3$ represents the phase shift, $i = 0,1,2$. The sample will be illuminated by the cosine structured patterns, and then the camera will capture the image of the sample.

The intensity value of light recorded by each pixel point of the camera (such as $A''(x_c, y_c)$) can be expressed as follows:

$$I(x_c, y_c; f_{x_o}, f_{y_o}; \phi_i)$$
$$= \eta \iint P(x_o, y_o; f_{x_o}, f_{y_o}; \phi_i) \times O(x_o, y_o) h_{\text{dect}}(x_c - \beta x_o, y_c - \beta y_o) dx_o dy_o \quad , \tag{2}$$

where $h_{\text{dect}}(x_o, y_o; x_c, y_c)$ represents the detection point spread function, $\beta$ represents the magnification of the imaging system.

With a three-step phase-shifting algorithm [34], the Fourier coefficient at the spatial-frequency $(f_{x_o}, f_{y_o})$ can be calculated as follows:

$$F_{(x_c, y_c)}(f_{x_o}, f_{y_o}) = \left[ 2I(x_c, y_c; f_{x_o}, f_{y_o}; 0) - I(x_c, y_c; f_{x_o}, f_{y_o}; 2\pi/3) - I(x_c, y_c; f_{x_o}, f_{y_o}; 4\pi/3) \right]$$
$$+ \sqrt{3} j \left[ I(x_c, y_c; f_{x_o}, f_{y_o}; 2\pi/3) - I(x_c, y_c; f_{x_o}, f_{y_o}; 4\pi/3) \right] \quad .$$
$$= 2b\eta \iint \left\{ \left[ O(x_o, y_o) \times h_{\text{dect}}(x_c - \beta x_o, y_c - \beta y_o) \right] \times e^{-j\varphi_0} e^{-j2\pi(f_{x_o} x_o + f_{y_o} y_o)} \right\} dx_o dy_o$$

(3)

The Fourier coefficient $F_{(x_c, y_c)}(-f_{x_o}, -f_{y_o})$ at the spatial-frequency $(-f_{x_o}, -f_{y_o})$ can also be calculated as follows by using the conjugate symmetric of the spectrum:

$$F_{(x_c, y_c)}(-f_{x_o}, -f_{y_o}) = \left[ 2I(x_c, y_c; f_{x_o}, f_{y_o}; 0) - I(x_c, y_c; f_{x_o}, f_{y_o}; 2\pi/3) - I(x_c, y_c; f_{x_o}, f_{y_o}; 4\pi/3) \right]$$
$$- \sqrt{3} j \left[ I(x_c, y_c; f_{x_o}, f_{y_o}; 2\pi/3) - I(x_c, y_c; f_{x_o}, f_{y_o}; 4\pi/3) \right] \quad . \tag{4}$$
$$= 2b\eta \iint \left\{ \left[ O(x_o, y_o) \times h_{\text{dect}}(x_c - \beta x_o, y_c - \beta y_o) \right] \times e^{j\varphi_0} e^{j2\pi(f_{x_o} x_o + f_{y_o} y_o)} \right\} dx_o dy_o$$

In addition, the Fourier coefficient $F_{(x_c, y_c)}(0,0)$ at the spatial frequency $(0,0)$ can be calculated as follows:

$$F_{(x_c, y_c)}(0,0) = \frac{1}{3} \left[ I(x_c, y_c; f_{x_o}, f_{y_o}; 0) + I(x_c, y_c; f_{x_o}, f_{y_o}; 2\pi/3) + I(x_c, y_c; f_{x_o}, f_{y_o}; 4\pi/3) \right]$$
$$= \frac{a}{3} \iint O(x_o, y_o) \times h_{\text{dect}}(x_c - \beta x_o, y_c - \beta y_o) dx_o dy_o$$

(5)

Based on the acquired Fourier coefficients $F_{(x_c, y_c)}(f_{x_o}, f_{y_o})$, $F_{(x_c, y_c)}(-f_{x_o}, -f_{y_o})$, and $F_{(x_c, y_c)}(0,0)$, a Fourier spectrum $\hat{I}_{(x_c, y_c)}(f_{x_o}, f_{y_o})$ is constructed. In this spectrum, all Fourier coefficients, except the points corresponding to the spatial frequencies $(f_{x_o}, f_{y_o})$, $(-f_{x_o}, -f_{y_o})$, and $(0,0)$, are zero. By applying the inverse Fourier transform to this Fourier spectrum, an image can be reconstructed:

$$
\begin{aligned}
I_{(x_c,y_c)}(x_o, y_o) &= \text{Real}\left\{\text{IFT}\left[\hat{I}_{(x_c,y_c)}(f_{x_o}, f_{y_o})\right]\right\} \\
&= 2\cos\varphi_0 \cos\left\{2\pi\left(f_{x_o}x_o + f_{y_o}y_o\right) + \arg\left[F_{(x_c,y_c)}(f_{x_o}, f_{y_o})\right]\right\}\left|F_{(x_c,y_c)}(f_{x_o}, f_{y_o})\right| \\
&\quad + 2\cos\varphi_0 \left|F_{(x_c,y_c)}(0,0)\right|
\end{aligned}
\quad (6)
$$

where $\text{IFT}[\cdot]$ represents the 2D inverse Fourier transform operator, $\text{Real}\{\cdot\}$ represents taking real part operator. $\left|F_{(x_c,y_c)}(f_{x_o}, f_{y_o})\right|$ and $\arg\left[F_{(x_c,y_c)}(f_{x_o}, f_{y_o})\right]$ represent the modulus and angle of the Fourier coefficient $\hat{I}_{(x_c,y_c)}(f_{x_o}, f_{y_o})$, respectively.

The image formulated in Eq. (6) is obtained by illuminating the sample with three structured light patterns sharing the same spatial frequency but differing in phase shift, followed by reconstruction using Fourier coefficients $F_{(x_c,y_c)}(f_{x_o}, f_{y_o})$, $F_{(x_c,y_c)}(-f_{x_o}, -f_{y_o})$, and $F_{(x_c,y_c)}(0,0)$. According to Fourier transform principle, this image corresponds to the frequency components of the dual image at $(f_{x_o}, f_{y_o})$, $(-f_{x_o}, -f_{y_o})$, and $(0,0)$ in the frequency domain. The intensity distribution of this image adheres to a cosine function whose spatial frequency matches that of structured illumination. When the spatial frequency of the structured light exceeds half of the objective lens's cutoff frequency, the frequency component surpassing the cutoff frequency can be effectively obtained. Furthermore, when the structured light frequency precisely aligns with the objective lens's cutoff frequency, the resulting frequency component attains an amplitude twice that of the cutoff frequency.

By extracting the value of the conjugated point from the image represented by Eq. (6), we can extract the frequency component beyond the cutoff frequency of the objective lens emitted from the conjugate object point, eliminate the signals emitted from the non-conjugate object point, and construct a super-resolution confocal image:

$$
\begin{aligned}
I_{\text{conf}}(x_c, y_c) &= \int I_{(x_c,y_c)}(x_o, y_o)\delta(x_o - x_{o\text{-}c}, y_o - y_{o\text{-}c})dx_s dy_s \\
&= I_{(x_c,y_c)}(x_{o\text{-}c}, y_{o\text{-}c}) \\
&= 2\cos\varphi_0 \cos\left\{2\pi\left(f_{x_o}x_{o\text{-}c} + f_{y_o}y_{o\text{-}c}\right) + \arg\left[F_{(x_c,y_c)}(f_{x_o}, f_{y_o})\right]\right\}\left|F_{(x_c,y_c)}(f_{x_o}, f_{y_o})\right| \\
&\quad + 2\cos\varphi_0 \left|F_{(x_c,y_c)}(0,0)\right|
\end{aligned}
\quad (7)
$$

where $(x_{o\text{-}c}, y_{o\text{-}c})$ represents the spatial coordinate of the object point conjugated to the camera pixel $(x_c, y_c)$.

While direct acquisition of coordinates of point $(x_{o\text{-}c}, y_{o\text{-}c})$ on the object plane is challenging, the conjugate relationship between the camera and the SLM can be calibrated to determine the coordinate of the point $(x_{s\text{-}c}, y_{s\text{-}c})$ on the SLM conjugated to camera pixel $(x_c, y_c)$. This calibration is feasible because the camera's photosensitive plane, the object plane, and the modulation plane of the SLM are optically conjugate, as illustrated in Fig. 1(b). The conjugate relationship between the SLM and the camera can be calibrated by using the method detailed in [28]. By leveraging the coordinate of the pixel $(x_{s\text{-}c}, y_{s\text{-}c})$, the signal emitted from the conjugate object point $(x_o, y_o)$ can be selectively extracted. Therefore, in practical implementation, when employing SR-CSIM with a single set of structured lights, we can extract the conjugate signal and construct the super-resolution confocal image using the following formula:

$$I_{\text{conf}}(x_c, y_c) = 2\cos\varphi_0 \cos\left\{2\pi\left(f_{x_o}\frac{x_{s\text{-}c}}{\alpha} + f_{y_o}\frac{y_{s\text{-}c}}{\alpha}\right) + \arg\left[F_{(x_c, y_c)}(f_{x_o}, f_{y_o})\right]\right\}\left|F_{(x_c, y_c)}(f_{x_o}, f_{y_o})\right| \\ + 2\cos\varphi_0 \left|F_{(x_c, y_c)}(0, 0)\right| \tag{8}$$

where $\alpha$ represents the magnification between the object plane and the modulation plane of the SLM.

Moreover, Eq. (8) demonstrates that the reconstruction of the dual image is not required in practical implementation. Instead, the coordinates of the conjugate points and the measured Fourier coefficients can be directly substituted into Eq. (8) to extract the conjugate signals. This approach substantially improves the efficiency of image reconstruction by bypassing computationally intensive intermediate steps.

The super-resolution confocal image reconstructed using a single set of unidirectional structured light patterns exhibits anisotropic resolution. This anisotropy arises because the resolution of the result is inherently dependent on the orientation of the structured illumination [35]. To achieve an anisotropic result, it is necessary to employ multiple sets of structured lights with spatial frequencies arranged in a ring of radius $R$ in the Fourier domain but varying orientations. This configuration enables the acquisition of multiple Fourier coefficients distributed uniformly across angular directions, ensuring consistent resolution enhancement in different spatial orientations.

When employing $K$ set of structured illumination patterns oriented in $K$ directions, the confocal image is mathematically expressed as follows:

$$I_{\text{conf}}(x_c, y_c) = 2\cos\varphi_0 \sum_{k=1}^{K} \cos\left\{2\pi\left(f_{x_o}^k \frac{x_{s\text{-}c}}{\alpha} + f_{y_o}^k \frac{y_{s\text{-}c}}{\alpha}\right) + \arg\left[\hat{I}_{(x_c, y_c)}\left(f_{x_o}^k, f_{y_o}^k\right)\right]\right\} \left|F_{(x_c, y_c)}\left(f_{x_o}^k, f_{y_o}^k\right)\right| \\ + 2\cos\varphi_0 \left|F_{(x_c, y_c)}(0, 0)\right|, \quad (9)$$

where $(x_{s\text{-}c}, y_{s\text{-}c})$ represents the coordinates of the pixel on the SLM optically conjugating to the camera pixel $(x_c, y_c)$. $k = 1, 2, \cdots, K$ is the index of measured Fourier coefficients. $\sqrt{\left(f_{x_o}^k\right)^2 + \left(f_{y_o}^k\right)^2} = R$, $R$ represents the radius of the frequency ring in the Fourier domain. As the number of structured light direction $K$ increases, the isotropy of the reconstructed image improves. However, an increase in $K$ also necessitates more measurements, which reduces the measurement speed. To achieve a practical balance, a minimum of $K = 3$ is recommended.

We can use Fig. 4 to summarize the image reconstruction method of the proposed SR-CSIM:

Step 1: Calibrate the pixel-wise conjugate relationship between the SLM pixel points (such as $A'$) and camera pixel points (such as $A''$).

Step 2: Illuminate the sample with high-frequency structured light patterns. For each camera pixel (such as $A''$), compute the Fourier coefficients using Eqs. (3-5).

Step 3: Apply Eq. (9) to extract the fluorescence signal emitted from the object point conjugated to the camera pixel (such as $A''$).

Step 4: Iterate Steps 2 and 3 on all camera pixels and reconstruct the super-resolution confocal image.

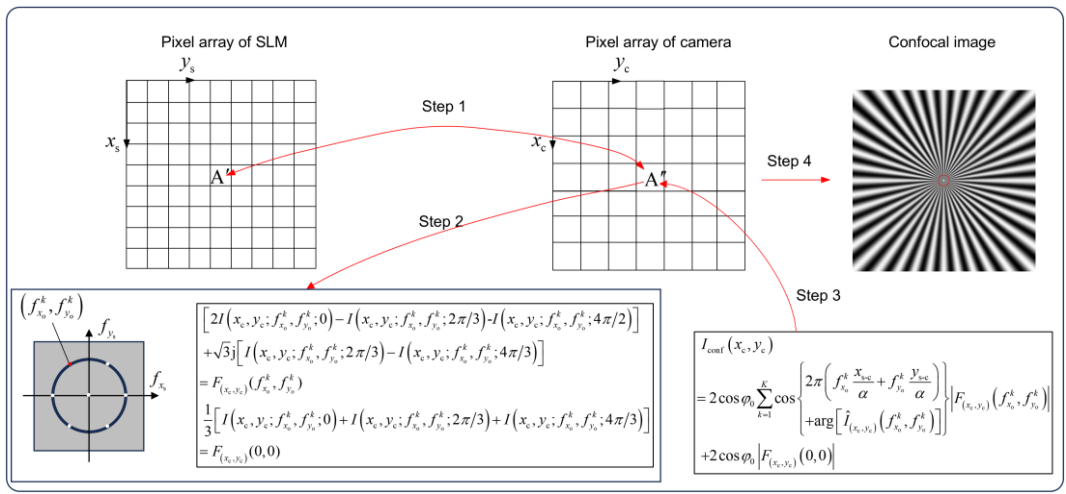

Fig. 4 Image reconstruction step of the proposed SR-CSIM.

# 3. Numerical simulation results

## 3.1 Result of the standard resolution target

To validate the effectiveness of the proposed method, we conducted simulation experiments using an objective lens with a numerical aperture (NA) of 1.49, $100\times$ magnification, and a fluorescence emission wavelength of 515 nm. Figure 5(a1) displays the simulated sample image, (size: $512\times512$ pixels; pixel size: $3.47\ \mu m$), while Fig. 5(a2) presents its Fourier spectrum. For comparison, Figure 5(b1) shows the image acquired using wide-field uniform illumination microscopy, with its spectrum (Fig. 5(b2)) bounded by the system's cutoff frequency (red circle, radius = 0.2 pixel$^{-1}$).

To achieve super-resolution imaging, the sample was illuminated with a single set of structured light patterns (spatial frequency: (0.2,0) pixel$^{-1}$) using three phase shifts ($0, 2\pi/3, 4\pi/3$). The camera captured three intensity images, and each camera pixel recorded three intensity values. Based on the three intensity values, the Fourier coefficient can be calculated by using Eqs. (3-5) to construct the spectrum (Fig. 5(c1)). Applying an inverse Fourier transform yielded the dual image (Fig. 5(c2)). From this dual image, the conjugate signal is extracted to construct the confocal super-resolution image (Fig. 5(c3)). Additionally, Figure 5(c4) presents the spectrum of this reconstructed image. Comparing Figs. 5(b1-b2) with Figs. 4(c3-c4), it is evident that the reconstruction result shown in Fig. 5(c3) exhibits improvement in resolution, and the radius of its spectrum is extended from 0.2 pixel$^{-1}$ to 0.4 pixel$^{-1}$, as shown in Fig. 5(c4). However, upon closer inspection, the resolution of the results in Figs. 5(c3-c4)) displays anisotropy. Specifically, the vertical resolution in Fig. 5(c3) has not improved, and the vertical spectrum in Fig. 5(c4) has not been extended. This discrepancy is primarily attributed to the use of a single set of structured lights with the same spatial frequency.

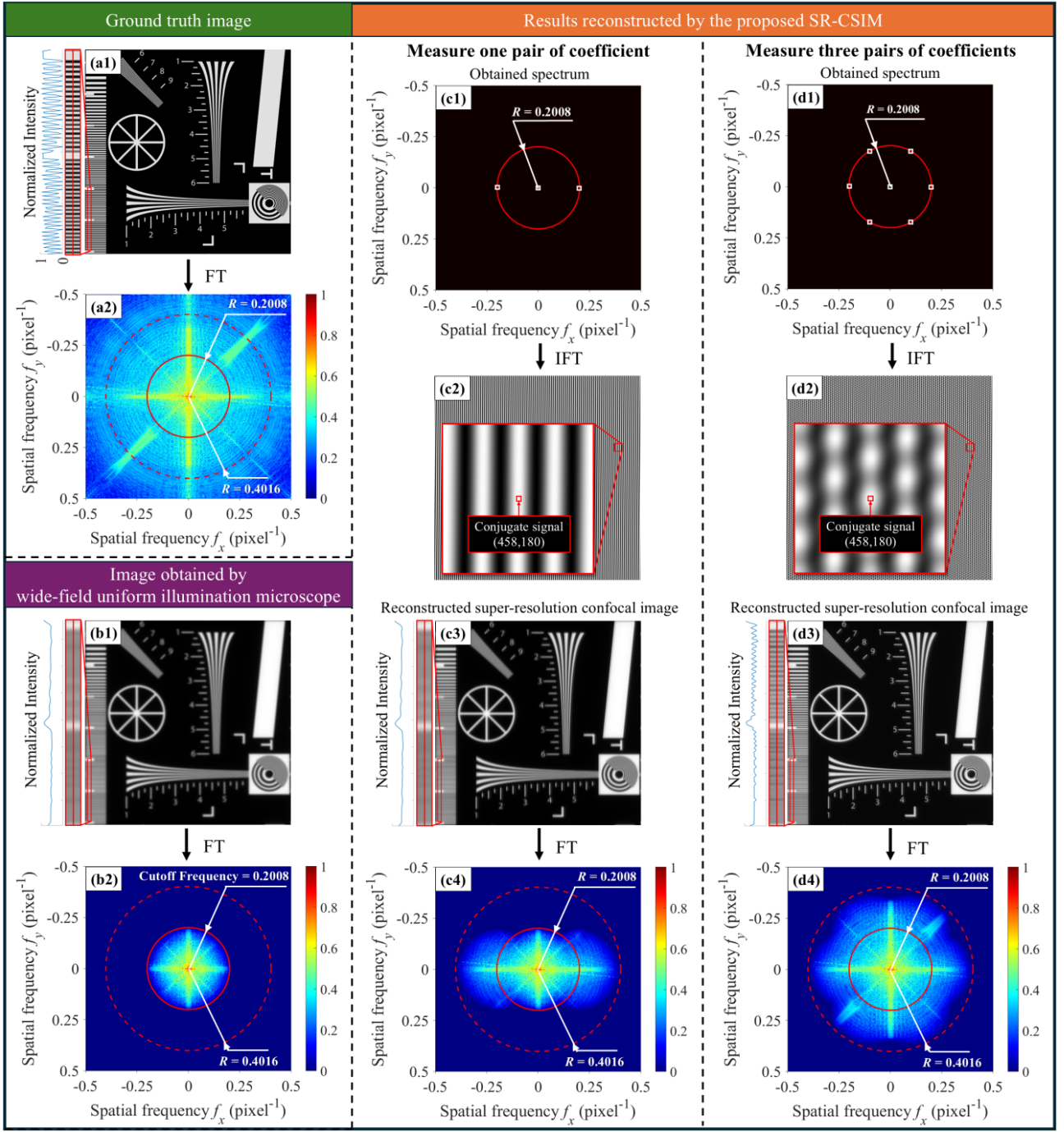

Fig. 5. Results of standard resolution target by using the proposed SR-CSIM. (a1-a2) Ground truth image and its spectrum; (b1-b2) image obtained by wide-field uniform illumination microscopy and its spectrum; the cutoff frequency of the objective lens is indicated by a red circle of a radius of 0.2 pixel$^{-1}$; (c1-c5) image reconstruction of the proposed SR-CSIM and using a set of structured light; (d1-d5) super-resolution confocal image reconstruction process by using the proposed SR-CSIM and using three sets of structured lights.

To achieve isotropic resolution in the reconstructed images, we employed three sets of structured light patterns with three different spatial frequencies to illuminate the sample. Each set of structured light patterns has the same spatial frequency but with three different phase shifts ($0, 2\pi/3, 4\pi/3$). The spatial frequencies of the nine structured light patterns are arranged in a ring of radius 0.2 pixel$^{-1}$ in the Fourier domain but with different directions. Following the methodology outlined in Subsection 2.2, three pairs of Fourier coefficients and a zero-frequency Fourier coefficient were computed and assembled a spectrum, as shown in Fig. 5(d1). Applying the inverse Fourier transform to this spectrum reconstructed the dual image (Fig. 5(d2)), from which conjugate signals were extracted to construct the confocal super-resolution image (Fig. 5(d3)). The resultant spectrum is shown in Fig. 5(d4)). By comparing these results (Figs. 4(d3-d4))) with those obtained using a single set of structured light (Figs. 4(c3-c4)), we observe that the resolution improvements are now isotropic. Specifically, both the vertical and horizontal resolutions in Fig. 5(d3) have been enhanced, and the spectrum in Fig. 5(d4) shows an extension in both vertical and horizontal directions. This demonstrates the effectiveness of employing three sets of structured lights from different orientations to address the previous anisotropy and achieve more uniform resolution enhancements across all directions.

**3.2 Result of Siemens star sample**

To further validate the robustness of the proposed SR-CSIM, we imaged a Siemens star resolution target. Figures 6(a1-a2) show the ground truth image of the sample and its corresponding Fourier spectrum. Figures 6(b1-b2) present the image acquired using wide-field fluorescence microscopy and its associated spectrum. These results demonstrate that wide-field fluorescence microscopy fails to resolve high-frequency information due to the diffraction limit. The diameter of the smallest resolvable circular feature in the wide-field fluorescence microscopy image was measured as D = 60.6 pixels. In contrast, in the ground truth image, the diameter of the smallest resolvable circular feature was measured as D = 23.2 pixels.

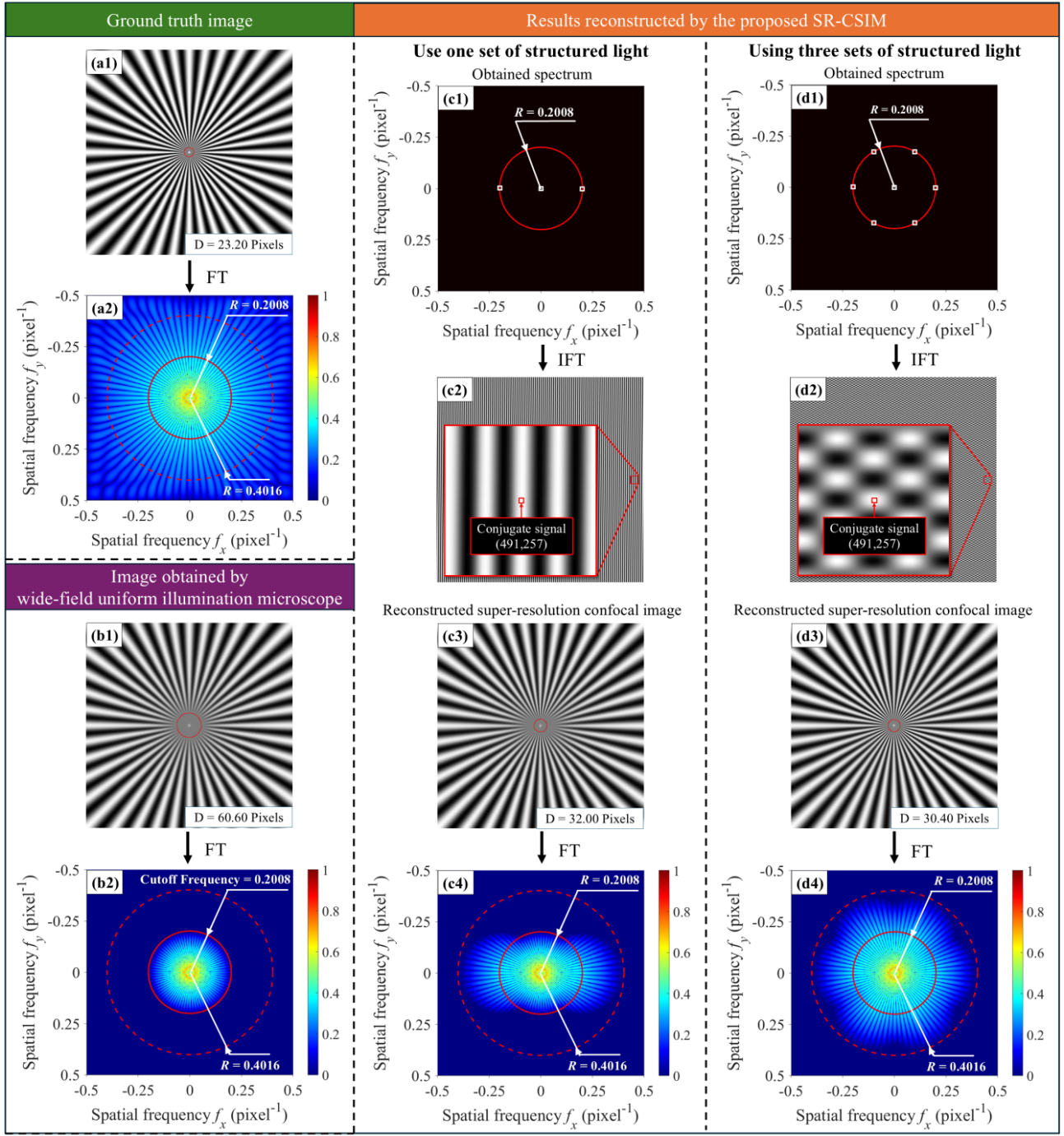

Fig. 6 Results of Siemens star sample. (a1-a2) Ground truth image and its spectrum; (b1-b2) image obtained by wide-field uniform illumination microscopy and its spectrum; the cutoff frequency of the objective lens is indicated by a red circle of a radius of 0.2 pixel$^{-1}$; (c1-c5) image reconstruction of the proposed SR-CSIM and using a set of structured light; (d1-d5) super-resolution confocal image reconstruction process by using the proposed SR-CSIM and using three sets of structured lights.

We then use the proposed SR-CSIM to image the Siemens star sample. Figures 6(c1-c4) illustrate the image reconstruction process and results using a single set of structured light patterns (identical spatial frequency with three initial phases), while Figs. 6(d1-d4) demonstrate the workflow and outcomes employing three sets of structured light patterns with distinct spatial frequencies. Figures 6(c1-c2) and 6(d1-d2) show the spectra obtained and the reconstructed dual images at the camera pixel point (491, 257) when using one set and three sets of structured - light patterns respectively. From these results, we observe the following findings:

- Using a single set of structured light patterns (same spatial frequency but three different initial phases), the diameter of the smallest resolvable circle improved from 60.6 pixels (wide-field microscopy, Fig. 6(b1)) to 32 pixels (Fig. 6(c3)). Frequency spectrum analysis (Fig. 6(c4)) revealed horizontal frequency extension to 0.4 pixel$^{-1}$, while the vertical frequency remained constrained by the system's original cutoff (0.2 pixel$^{-1}$), confirming direction-dependent resolution enhancement.

- With three sets of structured light patterns (different spatial frequencies), the minimum resolvable circle diameter was further reduced to 30.4 pixels (Fig. 6(d3)), corresponding to a 1.99-fold resolution improvement over wide-field microscopy. The radially symmetric spectrum (Fig. 6(d4)) exhibited uniform frequency extension to 0.4 pixel$^{-1}$ in all directions, validating isotropic resolution.

These findings demonstrate the effectiveness and robustness of the proposed method in achieving higher and more isotropic resolution in super-resolution confocal imaging.

### 3.3 Results of biological samples

To validate the effectiveness of our proposed method in addressing artifacts caused by fringe parameter estimation errors, we conducted simulation experiments. The experimental setup employed an objective lens with a magnification of $100\times$ and a numerical aperture of 1.4, operating at a fluorescence emission wavelength of 525 nm. The pixel size of the camera was $8\ \mu m$. The test sample, shown in Fig. 7(a1), was derived from open-source data provided by fairSIM [36] (https://www.fairsim.org/).

To emulate the effects of defocus and scattering noise, we introduced a direct current (DC) background into the fringe patterns, with the added DC background value being 0.1 and the fringe modulation being 0.1. This setup allowed us to test the robustness of our method under realistic imaging conditions. Figure 7 displays the results obtained by four different methods: the true image, wide-field fluorescence microscopy, SR-SIM, and our proposed SR-CSIM method. The SR-SIM reconstruction results were obtained using an open-source code provided in [37].

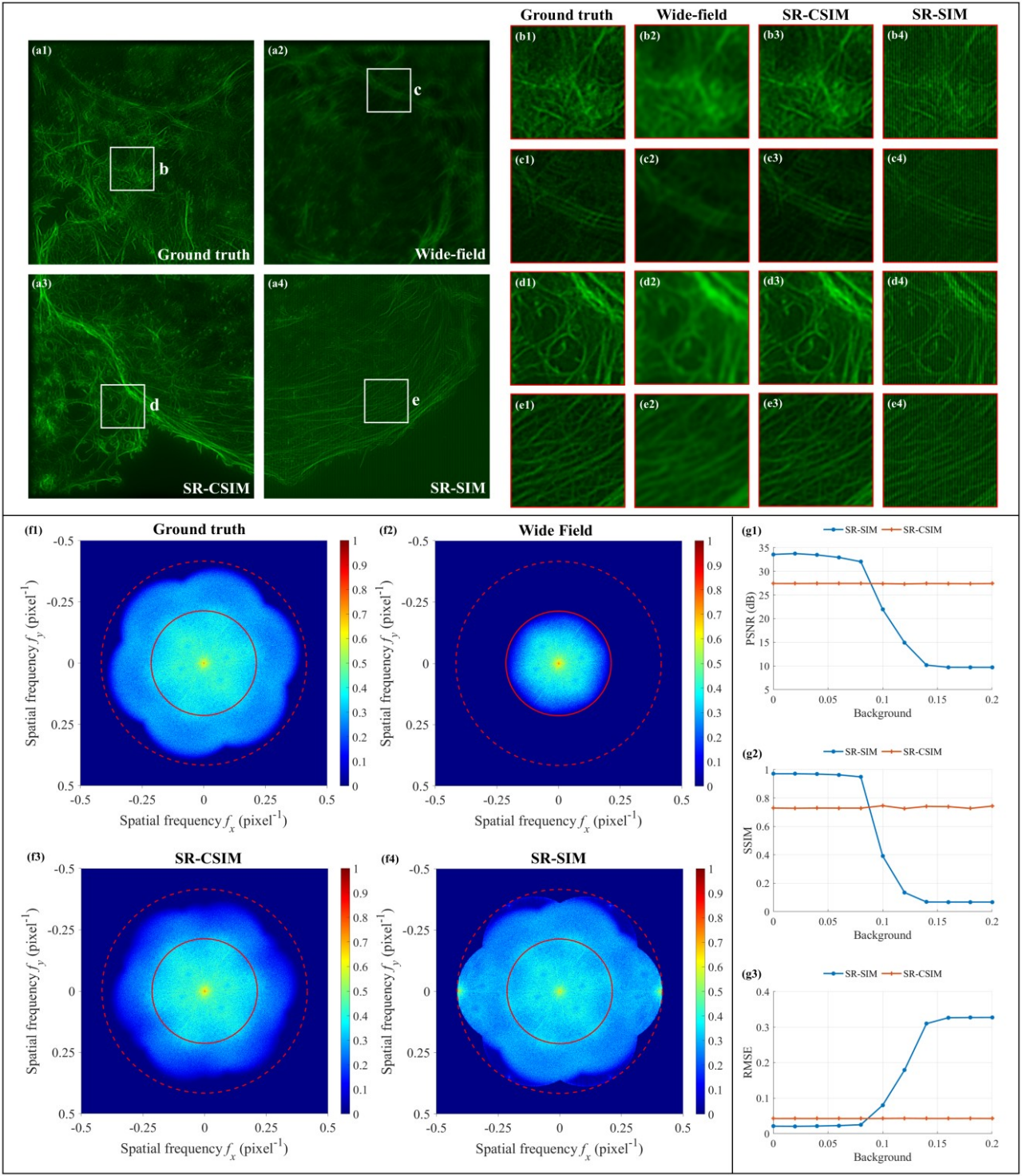

Figure 7. Performance comparison between the proposed SR-CSIM and the existing SR-SIM. (a1) Ground truth image; (a2) wide-field uniform illumination microscopy; (a3) SR-CSIM; (a4) SR-SIM. (b1-e1) Zoomed-in views of the ground truth image; (b2-e2) zoomed-in views of the result obtained by wide-field uniform illumination microscopy; (b3-e3) zoomed-in views of the reconstruction result using SR-CSIM; (b4-e4) zoomed-in views of the reconstruction result using SR-SIM; (f1-f4) Fourier spectrum of the images shown in Figs. (a1-a4); (g1-g3) comparison of the PSNR, SSIM, and RMSE of the results reconstructed by SR-CSIM and SR-SIM.

Table 1 provides a summary of the true values of fringe parameters (including spatial frequency ($k_x, k_y$), modulation, and initial phase ($\varphi_0, \varphi_1, \varphi_2$)) and the estimated values obtained by the SR-SIM method. The results reveal that the SR-SIM method introduces significant deviations in the estimation of fringe parameters compared to their true values. These discrepancies lead to erroneous spectral separation during the reconstruction process, ultimately resulting in artifacts in the final images, as shown in Figures 7(b4-e4). In contrast, our SR-CSIM method, which does not rely on fringe parameter estimation, produced artifact-free reconstruction results, as demonstrated in Figs. 7(b3-e3).

Table 1. Comparison between the parameters estimated by SR-SIM and the ground truth.

|  |  | $k_x$ | $k_y$ | Modulation | $\varphi_0$ | $\varphi_1$ | $\varphi_2$ |
|---|---|---|---|---|---|---|---|
| Ground | $K_A$ | -0.1025 | -0.1748 | 0.1 | 0° | 120° | −120° |
| truth | $K_B$ | 0.1025 | -0.1748 | 0.1 | 0° | 120° | −120° |
| parameters | $K_C$ | -0.2031 | 0 | 0.1 | 0° | 120° | −120° |
| Parameters | $K_A$ | -0.1025 | -0.1748 | 0.2148 | 0.0752° | 119.9201° | −119.8556° |
| estimated | $K_B$ | 0.1025 | -0.1748 | 0.1983 | −0.0036° | 119.7052° | −120.0203° |
| by SR-SIM | $K_C$ | -0.2036 | 0 | 0.2549 | −15.5486° | 122.2047° | −154.2725° |

To further quantify the performance of the proposed method, we analyzed the frequency spectra of the reconstructed images. As shown in Figure 7(f1-f4), spectral comparisons reveal that SR-SIM erroneously amplifies high-frequency components at boundary regions (Fig. 7(f4)), deviating from the ground-truth spectrum (Fig. 7(f1)). In contrast, the SR-CSIM method (Fig. 7(f3)) maintains spectral consistency without such distortions.

Finally, Figure 7(g1-g3) shows the variation curves of PSNR, SSIM, and RMSE for the reconstructed results of SR-CSIM and SR-SIM as the DC background value increased. The results demonstrate that as the DC background increased, the performance of SR-SIM deteriorated significantly (the PSNR and SSIM values decreased while the RMSE value increased). In contrast, the PSNR, SSIM, and RMSE values of the SR-CSIM reconstructed results remained stable. This outcome substantiates the robustness of our proposed SR-CSIM method in mitigating artifacts caused by fringe parameter estimation errors.

It should be noted that when the added background noise is below 0.08, the PSNR, SSIM, and RMSE metrics of SR-SIM are slightly superior to those of SR-CSIM. This is primarily because SR-SIM employs deconvolution in the reconstruction process, whereas SR-CSIM does not utilize deconvolution. However, as the background noise increases, the advantages of SR-CSIM in maintaining stable and artifact-free reconstruction results become evident, demonstrating its superior performance in challenging imaging conditions. These findings collectively validate the feasibility and effectiveness of our proposed approach.

## 4. Discussion

In the process of image reconstruction, the inclusion of the zero-frequency Fourier coefficient enhances the brightness of the confocal image. However, this comes at the cost of reduced contrast due to the contribution of out-of-focus signals embedded within this coefficient. To address this trade-off, we propose weighting the zero-frequency Fourier coefficient using an empirically determined factor. This approach allows for the optimization of the balance between brightness and contrast based on specific application requirements.

When employing multiple sets of structured illumination with distinct spatial frequencies, a zero-frequency Fourier coefficient is calculated for each set. To minimize the impact of measurement errors, the final zero-frequency coefficient is obtained as the arithmetic mean of all individual coefficients. This averaging process reduces the influence of noise and ensures a more reliable reconstruction of the spectra components.

The SR-CSIM requires pre-calibration of the conjugate relationship between the SLM and the camera. This calibration step ensures accurate extraction of conjugate signals. As described in [28], the calibration process is straightforward and remains stable provided the optical system remains unchanged (i.e., no positional variations in the optical components). Once calibrated, this system does not require recalibration for repeated measurements under stable experimental conditions, thereby enhancing workflow efficiency and practicality.

In comparison to conventional SR-SIM, the proposed SR-CSIM addresses the artifact limitations associated with structured illumination parameter estimation errors. By eliminating the need for fringe parameter estimation, our method significantly improves imaging accuracy and reliability. This technological breakthrough not only enhances the ability to observe fine structural details in biomedical imaging but also extends its potential for nanoscale structural analysis in materials science. Furthermore, the broad applicability of this method across complex biological systems and materials science is poised to drive advancements in health and related industries.

## 5. Conclusion

This paper proposes the SR-CSIM. The physical imaging model of SR-CSIM is based on the principle of confocal imaging. Experimental results demonstrate that when the spatial frequency of the structured light matches the cutoff frequency of the objective lens, the resolution of the reconstructed images by SR-CSIM is enhanced by nearly a factor of two compared with that of conventional wide-field fluorescence microscopy. Importantly, compared to SR-SIM, SR-CSIM does not need to estimate the parameter of structure light, thus effectively avoiding artifacts induced by the parameter estimation error. This feature enables SR-CSIM to expand the application scope of structured illumination microscopy.